\DocumentMetadata{}
\documentclass[sigconf]{acmart}
\usepackage{booktabs} 
\usepackage{enumitem}
\usepackage{balance}
\usepackage[utf8]{inputenc}
\usepackage{multirow}
\usepackage{graphicx}
\usepackage{microtype}
\usepackage{subfigure}
\usepackage{bm}
\usepackage{amsmath}
\usepackage{amsfonts}
\usepackage{graphicx}

\newcommand{\squishlist}{
 \begin{list}{$\bullet$}
  { \setlength{\itemsep}{0pt}
     \setlength{\parsep}{3pt}
     \setlength{\topsep}{3pt}
     \setlength{\partopsep}{0pt}
     \setlength{\leftmargin}{1.5em}
     \setlength{\labelwidth}{1em}
     \setlength{\labelsep}{0.5em} } }

\newcommand{\squishlisttwo}{
 \begin{list}{$\bullet$}
  { \setlength{\itemsep}{0pt}
     \setlength{\parsep}{0pt}
    \setlength{\topsep}{0pt}
    \setlength{\partopsep}{0pt}
\setlength{\leftmargin}{2em}
\setlength{\labelwidth}{1.5em}
\setlength{\labelsep}{0.5em} } }

\newcommand{\squishend}{
\end{list}  }

\AtBeginDocument{%
  \providecommand\BibTeX{{%
    \normalfont B\kern-0.5em{\scshape i\kern-0.25em b}\kern-0.8em\TeX}}}

\setcopyright{acmcopyright}
\copyrightyear{2018}
\acmYear{2018}
\acmDOI{10.1145/1122445.1122456}

\acmConference[Woodstock '18]{Woodstock '18: ACM Symposium on Neural
  Gaze Detection}{June 03--05, 2018}{Woodstock, NY}
\acmBooktitle{Woodstock '18: ACM Symposium on Neural Gaze Detection,
  June 03--05, 2018, Woodstock, NY}
\acmPrice{15.00}
\acmISBN{978-1-4503-XXXX-X/18/06}

\begin{document}

\title{Serendipitous Recommendation with Multimodal LLM}



\author{Haoting Wang$^1$, Jianling Wang$^1$, Hao Li$^2$, Fangjun Yi$^2$, Mengyu Fu$^2$, Youwei Zhang$^2$, Yifan Liu$^2$, \\ Liang Liu$^2$, Minmin Chen$^1$, Ed H. Chi$^1$, Lichan Hong$^1$ and Haokai Lu$^1$}
\affiliation{%
  \institution{$^1$ Google DeepMind \quad 
 $^2$YouTube}
}
\email{{haotingwang,jianlingw,hlg,fangjunyi,mengyufu,youweiz,yifanliu,liangliu,minminc,edchi,lichan,haokai}@google.com}













\begin{abstract}
Conventional recommendation systems succeed in identifying relevant content but often fail to provide users with surprising or novel items. Multimodal Large Language Models (MLLMs) possess the world knowledge and multimodal understanding needed for serendipity, but their integration into billion-item-scale platforms presents significant challenges. In this paper, we propose a novel hierarchical framework where fine-tuned MLLMs provide high-level guidance to conventional recommendation models, steering them towards more serendipitous suggestions. This approach leverages MLLM strengths in understanding multimodal content and user interests while retaining the efficiency of traditional models for item-level recommendation. This mitigates the complexity of applying MLLMs directly to vast action spaces. We also demonstrate a chain-of-thought strategy enabling MLLMs to discover novel user interests by first understanding video content and then identifying relevant yet unexplored interest clusters. Through live experiments within a commercial short-form video platform serving billions of users, we show that our MLLM-powered approach significantly improves both recommendation serendipity and user satisfaction.
\end{abstract}



\renewcommand{\shortauthors}{Haoting Wang et al.}
\maketitle

\section{Introduction}
Ubiquitous across diverse online platforms,  recommendation systems \cite{li2023text,li2023gpt4rec,hou2023large,geng2022recommendation} play a critical role in helping users navigate the vast and ever-growing content available online, connecting them with relevant items based on their past behavior and preferences.
However, while traditional approaches excel at reinforcing past interests, an over-emphasis on narrow relevance can easily confine users within their established interest areas, limiting exposure to new ideas and experiences. Consequently, there is an exciting and increasingly critical opportunity to enhance these systems beyond mere relevance, aiming to improve overall user value by introducing users to a wider world of discovery \cite{chen2021exploration,wu2024result}.

This pursuit of broader user horizons motivates the development of \textit{serendipitous recommendation systems}. Unlike purely relevance-focused systems, they aim to suggest items that are not only relevant but also novel, surprising, and unexpected, holding the key to sparking curiosity, fostering exploration, and ultimately enriching the user experience \cite{ziarani2021serendipity,chen2021values}. The goal is to strike a delicate balance: reliably recommending familiar items that align with established user preferences while strategically introducing novel items that encourage exploration and broaden perspectives.

However, achieving meaningful serendipity is inherently challenging. Core feedback mechanisms in traditional recommendation systems often create self-reinforcing loops, heavily favouring items similar to users' past interactions. Recent advancements have explored the potential of Large Language Models (LLMs) to mitigate this issue by leveraging their vast world knowledge and reasoning capabilities to infer user interest beyond their behavioral signals \cite{wang2024large,wang2024llms}. While LLMs offer promising avenues for injecting novelty and breaking the feedback loop, how to bridge the gap between their general world knowledge and the specific nuances of domain-specific items and user interactions is a key difficulty.

Furthermore, the rise of multimodal content, particularly in domains like short-form video, reveals the limitations of text-only analysis. Understanding user preferences and item characteristics requires engaging with crucial visual elements, presenting both new challenges and opportunities. Multimodal LLMs (MLLMs) \cite{achiam2023gpt,team2023gemini,driess2023palm}, capable of processing and integrating information from diverse inputs like text, images, and potentially audio and video, present an exciting frontier for richer content understanding and more nuanced user behavior alignment. However, harnessing these powerful but computationally intensive models within the constraints of large-scale recommendation systems introduces significant engineering complexities, requiring efficient inference and effective management of the vast action space of recommendations.

\begin{figure}
    \centering
    \includegraphics[width=0.45\textwidth]{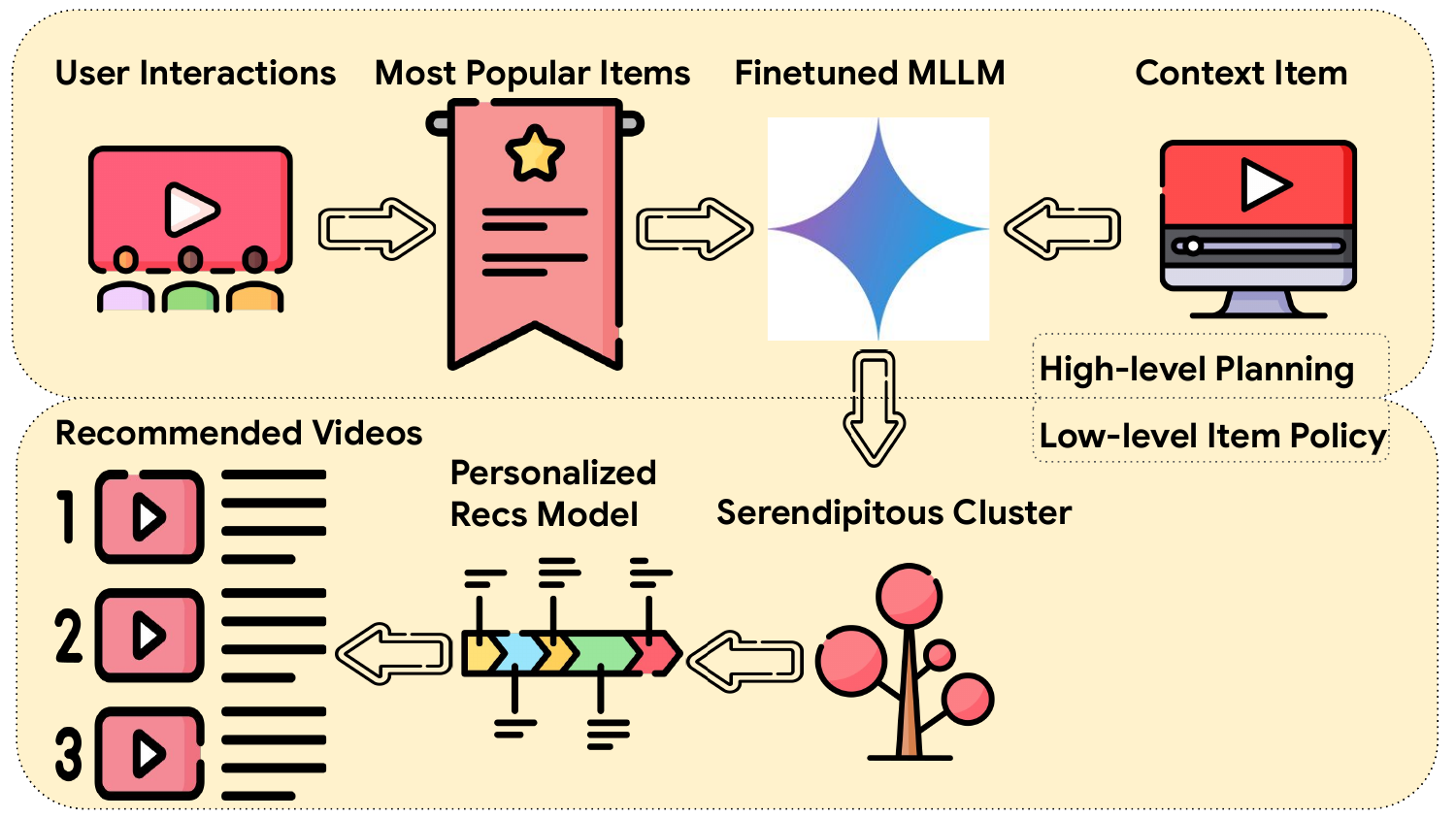}
    \caption{Large-scale recommendation through hierarchical planning framework With MLLM.}
    \label{fig:diagram}
\end{figure}

To tackle the aforementioned challenges of fostering serendipity, leveraging multimodal understanding, and ensuring practical scalability, we propose a novel paradigm that integrates MLLMs with classic recommendation models, as illustrated in Figure \ref{fig:diagram}. Our approach utilizes the sophisticated understanding capabilities of MLLMs to enhance item representation and user behavior modelling, specifically aiming to identify and promote content likely to induce serendipitous discovery. This enhanced understanding is then seamlessly integrated into a robust, industrial-scale recommendation framework capable of handling billions of users. This work marks a pioneering effort to leverage the power of multimodal large language models specifically for increasing recommendation serendipity within a practical, large-scale industrial setting. Specifically, our model and contribution possess the following unique characteristics: (i) We provide a scalable solution capable of delivering personalized serendipitous recommendations sensitive to the most recent user interactions. (ii) We explore and implement a feasible pathway for utilizing MLLMs for deep video understanding and incorporating these insights effectively into an industrial-scale recommendation pipeline. (iii) Through live experiments on a commercial short-form video recommendation platform serving billions of users, we demonstrate that our MLLM-powered serendipitous recommendation pipeline not only significantly improves the novelty and diversity of recommendations but also leads to measurable improvements in user satisfaction metrics.

\smallskip
\noindent\textbf{Related Work.} 
The recent advancement in LLMs has opened exciting avenues for enhancing recommendation systems. 
Some studies explore using LLMs directly for generating recommendations, by directly generating language description \cite{bao2023tallrec,dai2023uncovering,geng2023vip5,hou2023large,li2023gpt4rec,liu2023chatgpt}, or through token-based methods \cite{zhaiactions,singh2024better}.
Others focus on augmenting traditional recommendation models with LLM-powered feature engineering \cite{hou2023learning,yu2021tiny} or enriched user/item representations \cite{li2023taggpt,liu2023first,xi2023towards}, which utilize LLMs to generate embeddings that are then integrated into conventional recommendation systems. 
Specifically, to handle multimodal information, NoteLLM \cite{zhang2024notellm} and Rec-GPT4V\cite{liu2024rec} propose to integrate and process
diverse data types such as images, text, and video to generate embeddings.
To address latency for large-scale recommendation systems,  Wang et al. \cite{wang2024large} use LLMs as data augmenters for conventional recommendation systems during training, to improve model performance without additional serving cost. And in \cite{wang2024llms}, they proposed to incorporating LLM-generated content to break the feedback loop with a hierarchical planning structure and utilizing LLMs to provide high-level language policy guidance. This work represents a novel attempt to leverage MLLMs for multimodal understanding and enhancing serendipity in a practical, industrial-scale recommendation system.

\section{Preliminaries}
\label{sec:preliminary}
Industrial-scale recommendation systems need to handle an overwhelmingly large action space, making it inefficient for LLMs to directly identify the next best item. To address this, we adopt the \textbf{LLM-based Hierarchical Recommendation} framework proposed in \cite{wang2024llms}. Specifically, it leverages LLMs for high-level planning at the interest cluster level, effectively narrowing down the search space. These clusters are then fed into a conventional recommendation model for real-time item-level recommendations. To facilitate this, we employ \textit{hierarchical tree-structure clustering} methods \cite{chang2024cluster} to obtain high-quality item clusters. This involves creating equal-sized and traffic-weighted clusters based on item topics, resulting in a 4-level tree structure.  An item is associated with the tree nodes at different levels, with higher-level clusters representing broader topics and lower-level clusters representing more specific ones. Each cluster is assigned a unique natural language description to encapsulate its topical focus.

\smallskip
\noindent\textbf{Serendipitous User Experience}  Beyond scalability, we target a specific problem in contextual recommendation: given an item a user engaged with, what subsequent items optimize the user experience? While collaborative filtering and transformer-based models are effective at finding similar items, these solutions can easily introduce a strong feedback loop, potentially leading to stale user experiences. Our hypothesis is that serendipitous recommendations, defined as items that are relevant yet also unexpected or surprising \cite{chen2021values}, can lead to a better user experience than relying solely on similarity. 

Using the established tree-structure clusters, we can formalize this concept: We define a context item  $v_i$  and a recommended item  $n_j$ as similar if they fall into the same cluster node at tree level $l$; Conversely, they are considered serendipitous if they fall into different cluster nodes at level $l$ but the same cluster at a broader level  $l-\delta$. On an industrial video recommendation platform, we aggregated user satisfaction rates for these two types of recommended items, and observed significantly higher satisfaction with serendipitous recommendations, validating our hypothesis.

Based on this analysis, our goal is to find high quality (context item, serendipitous recommended clusters) pairs from user logs that represent positive serendipitous user experiences. We aim to use these examples to fine-tune an MLLM, enabling it to leverage its world knowledge to identify and suggest novel, serendipitous clusters beyond those already prevalent in the system logs.

\smallskip
\noindent\textbf{MLLM Fine-Tuning for Serendipitous Experience}
We propose to inject domain knowledge, and desired serendipitous behavior through fine-tuning. For each context video $v_i$, we collect the next video $n_i$ user interacted with, along with the users satisfaction rate $p_{ij}$ on that next video. We only keep the $(v_i, n_j)$ pair if it is serendipitous, i.e., $v_i$ and $n_j$ fall into different cluster nodes on tree level $l$ but same cluster on tree level $l-\delta$. We further use $p_{ij}$ to select serendipitous pairs with highest user satisfaction.

This hierarchical recommendation framework combines the strengths of both MLLMs and conventional recommendation models. By leveraging fine-tuned MLLMs capable of \textbf{controlled generation}~\cite{wang2024llms}, it first identifies a serendipitous cluster for the context item. The MLLM generates a cluster description, which is then translated into a corresponding cluster ID. At the low level, this cluster ID is used to restrict a conventional transformer-based sequence model to efficiently retrieve items from within the selected cluster. The framework leverages LLM's generalization and reasoning capabilities to discover serendipitous interest, and yet it exploits a domain-specific sequence model to improve personalization and handle item dynamics.



\section{Serendipitous Recommendation with Multimodal LLM}

This section details our proposed method, focusing on how it addresses two key challenges: (1) How to effectively represent items by integrating information from multiple modalities, especially complex visual data; and (2) How to maintain scalability while handling a vast number of items and their associated modalities.
\label{sec:finetune}

\subsection{Item level understanding with MLLMs}

Prior works \cite{wang2024llms,wang2024large} primarily rely on textual representations of items, such as video titles, descriptions, or transcripts, to understand their content for improving recommendation performance. However, we argue that directly incorporating video itself allows for a more nuanced and detailed understanding, particularly for tasks like serendipitous recommendation where subtle thematic connections are important. We utilize fine-tuned MLLMs (Section \ref{sec:preliminary}) for this purpose. A central question is how best to represent items that integrate both visual and textual content within the recommendation pipeline. To evaluate different approaches, we designed offline metrics:
\squishlist
\item Match rate: the percentage of LLM outputs that exactly matches any cluster description, to measure model's understanding of domain knowledge.
\item Recall: the percentage of LLM output that exactly matches the label cluster description, to measure model's ability to generate high-quality serendipitous journeys.
\squishend



We conducted a series of offline experiments exploring different input modalities (textual: titles, cluster descriptions; visual: thumbnails, sampled frames) and prompting strategies to answer key questions regarding item representation:

\begin{figure}
    \centering
    \includegraphics[width=0.45\textwidth]{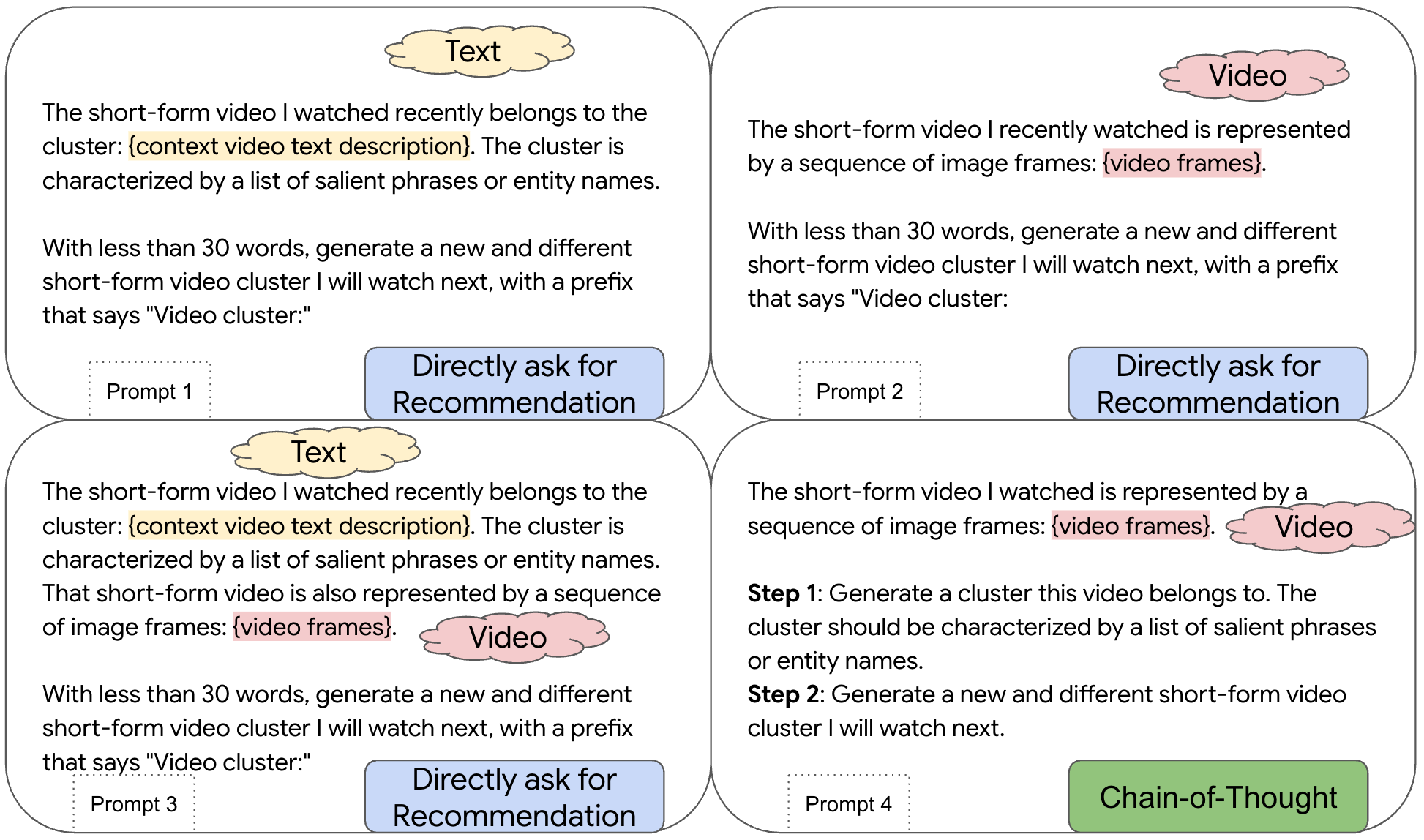}
    \caption{Different Types of Prompt. We compare prompts across two dimensions: input modality and prompting strategy. Modalities include text-only (Prompt 1), video-only (Prompt 2 \& 4), and multimodal (Prompt 3). We compare a direct-prompting strategy (Prompts 1-3) against chain-of-thought reasoning (Prompt 4).}
    \label{fig:prompt definition}
\end{figure}



\begin{figure}[h!]
    \centering
    \includegraphics[width=0.43\textwidth]{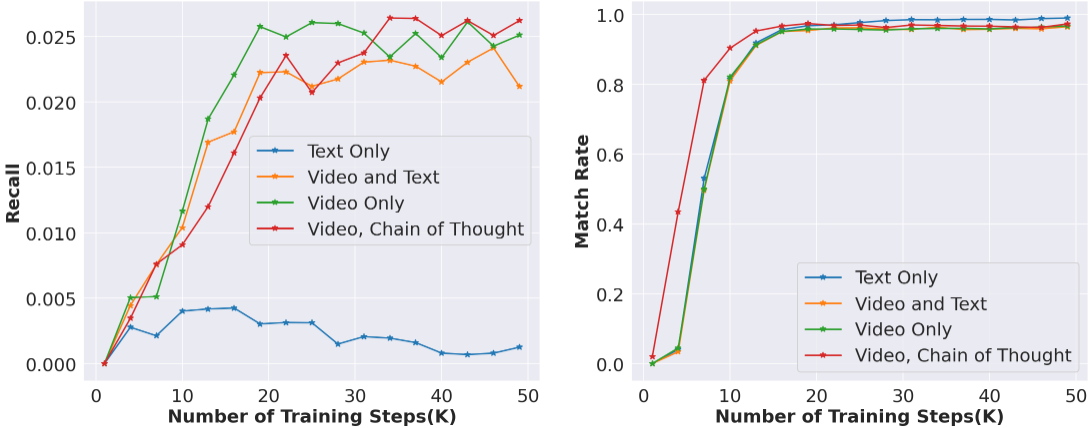}
    \caption{Offline Metrics: Different Types of Prompt. Models using video input significantly outperform the model using only text input. Among the video-based models, chain-of-thought prompting delivers the best performance.}
    \label{fig:eval_prompt_version}
\end{figure}



\smallskip
\noindent\textbf{How effective is Chain-of-Thought for representing visual content?}
We compared different prompting approaches and input types, as shown in Figure \ref{fig:prompt definition}. Figure \ref{fig:eval_prompt_version} presents the performance during fine-tuning (batch size 32, 4 uniformly sampled frames for video input, cluster text description for text input). The results first confirm the value of visual data: adding video input significantly boosts Recall compared to text-only approaches.

However, effective integration is crucial. A naive combination of video and text using direct prompting (Video and Text) actually results in a significant drop in Recall compared to using video alone. This highlights the inadequacy of simple concatenation and underscores the need for structured reasoning when processing multimodal inputs, particularly visual information.

Therefore, we propose and evaluate Chain-of-Thought (CoT) prompting \cite{wei2022chain} with fine-tuned MLLMs, specifically to generate structured, interpretable representations of visual item content. CoT prompts the MLLM to break down the visual analysis into intermediate reasoning steps before outputting a final representation or recommendation, potentially capturing more subtle cues than direct generation. Indeed, our experiments show that employing CoT with video input (Video, CoT) leads to strong Recall performance (achieving comparable peak performance to Video Only) and notably accelerates the convergence of the Match Rate. This faster convergence suggests that the CoT strategy effectively guides the MLLM to better understand visual content and acquire domain knowledge more rapidly.

\begin{figure}[h!]
    \centering
    \includegraphics[width=0.43\textwidth]{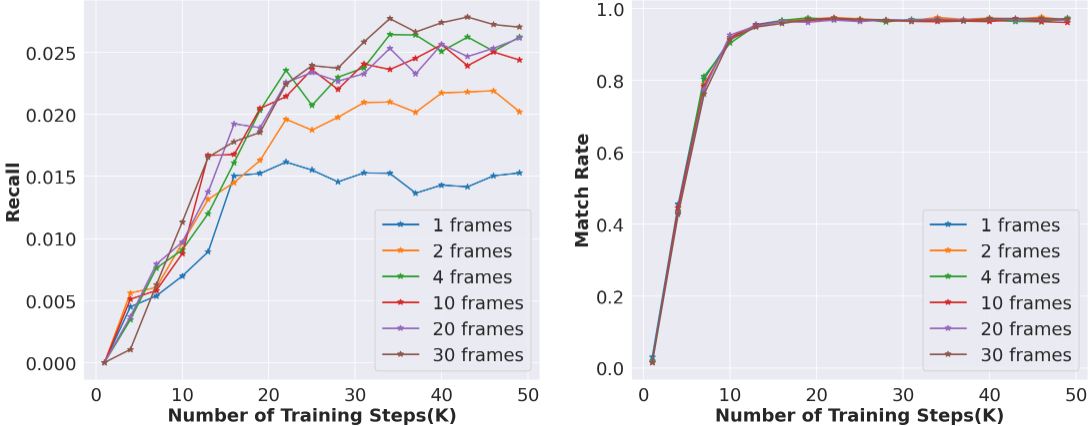}
    \caption{Offline Metrics Different Number of Frames. Left: Recall benefits from more frames, but with diminishing returns beyond 4 frames. Right: match rate is insensitive to the frame count and saturates quickly.}
    \label{fig:NumFrames}
\end{figure}

\smallskip
\noindent\textbf{What's the optimal strategy for sampling video frames?}
Having established the value of visual input, particularly when processed thoughtfully (e.g., via CoT), we examined frame sampling. Figure \ref{fig:NumFrames} shows the impact of varying the number of uniformly sampled frames (using CoT prompting, batch size 32). Increasing frames up to 4 improves recall stability and convergence speed. Beyond 4 frames, the improvement diminishes for the short videos (< 1 min) analyzed, suggesting a trade-off between information gain and computational cost.

\begin{figure}[h!]
    \centering
    \includegraphics[width=0.43\textwidth]{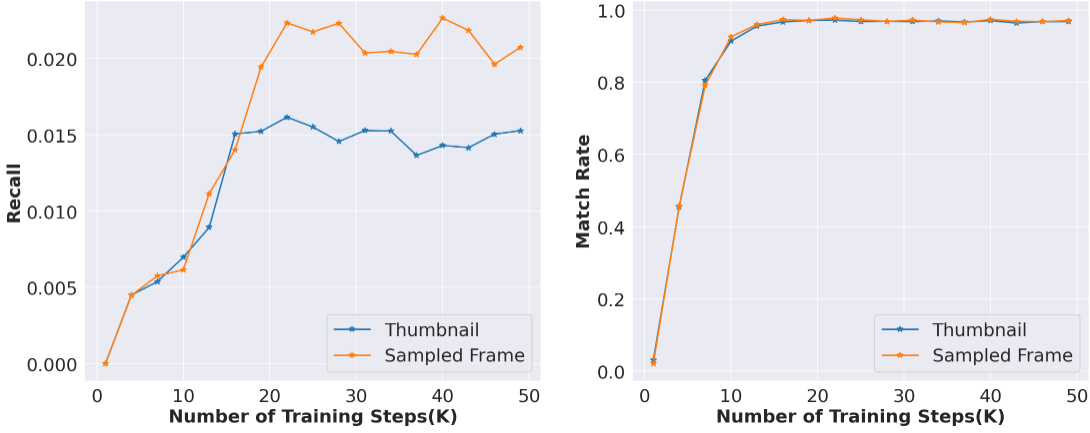}
    \caption{Offline Metrics Thumbnail VS 1 Sampled Frame. Training with thumbnails results in higher recall compared to training with sampled frame.}
    \label{fig:thumbanil_vs_frames}
\end{figure}

\smallskip
\noindent\textbf{Are thumbnails or sampled frames better visual input?}
Figure \ref{fig:thumbanil_vs_frames} compares the performance of using thumbnail image versus uniformly sampled frame as video input, with batch size of 32 and chain-of-thought prompting style. The results show that using thumbnails leads to significantly higher recall after convergence. This indicates that thumbnails, often curated to be representative, serve as a more potent visual summary for the MLLM than a single random frame, especially when processed using our CoT approach.



\subsection{Scalability} We tackle contextual recommendations by focusing on item representation, which presents a significant challenge due to the sheer scale and dynamic nature of industrial recommendation platforms. With billions of items, and new ones constantly added or removed, efficiently processing lengthy visual and textual representations becomes crucial to achieve comprehensive coverage and fast inference, all while minimizing resource consumption.

To address this issue, we implemented a scalable serving infrastructure illustrated in Figure \ref{fig:diagram}.  By limiting the item corpus to the most popular videos, we are able to achieve 80\% impression coverage and finish the inference within 12 hours. This inference job runs regularly to ensure continuous coverage on impressions. We further boosted efficiency by adopting an incremental update approach. Since the most popular videos don't change drastically between runs, we only process new additions, speeding up inference by 5x on average.

\section{Live Experiments}
\label{sec:live_exp}
\subsection{Experiment Setup}
To examine the proposed method, we conducted live experiments on a commercial short-form video recommendation platform that serves billions of users. We used Gemini 1.5 \cite{team2024gemini} as the LLM for video understanding and reasoning. We selected equal amount, non-overlapping segments of user traffic for control and experiment arms, and ran the experiments for over 30 days. We selected the chain-of-thought style prompt with 4 uniformly sampled frames. The fine-tuning process and serving infrastructure could be easily adapted to other LLMs and types of prompts.

\smallskip
\noindent\textbf{Diversified Data Curation.}  
We group $(v_i, n_j)$ pairs by next video $n_j$'s cluster $C(n_j)$, resulting in $(v_i, C(n_j))$ pairs. We also aggregate the user satisfaction rate on the next video' cluster $C(n_j)$ for each context video $v_i$, getting $p_{ic(n_j)}$. Then for each $C(n_j)$, we select top 10 context videos with highest $p_{ic(n_j)}$ into the training data. This approach ensures that the training data not only represents improved user value, but also has an equal portion for different clusters. The training data included all the cluster descriptions, so the model can be finetuned to achieve controlled generation.

\smallskip
\noindent\textbf{Baseline.} 
We compared the performances of the proposed methods with different types of baseline models that are currently in production: \textbf{(1) Exploitation-oriented model}: the transformer-based model \cite{chen2019top} trained on users consumption history sequence and positive feedback. \textbf{(2) Exploration-oriented models}, including a transformer-based model that prioritize items that are more similar to context item rather than user's long consumption history; and a neural linear bandit-based model to predict next novel cluster. Our online experiment shows that our proposed method shows better novelty and quality compared to the existing models.

\begin{figure}[t]
    \centering
    \includegraphics[width=0.28\textwidth]{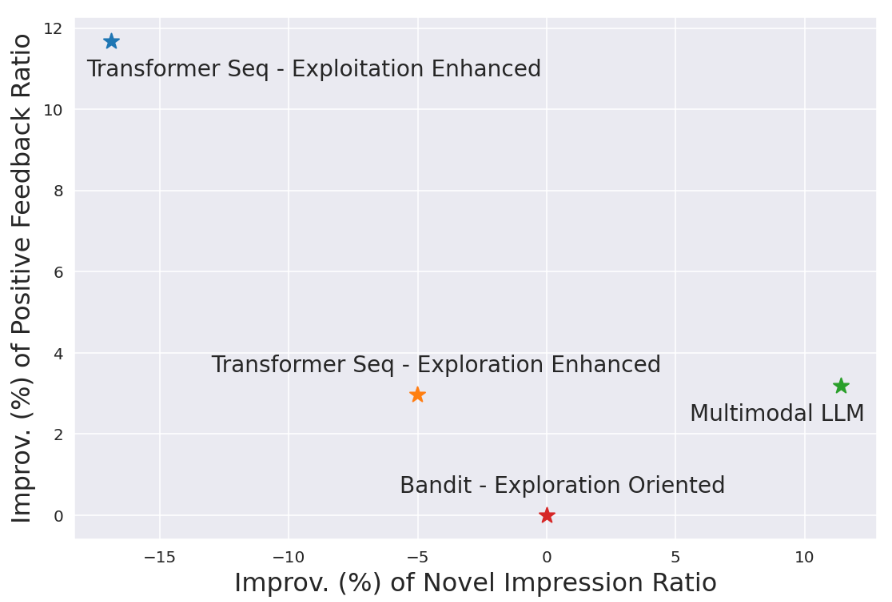}
    \caption{Novelty and Quality Comparison. The proposed serendipitous recommendation with MLLM achieves the highest novelty gain.}
    \label{fig:novelty}
\end{figure}

\subsection{Results and Analysis}


\noindent\textbf{Serendipity.}
In Figure \ref{fig:novelty}, we compared the proposed method with various baseline models currently in production. Using the exploitation-oriented bandit model as a reference, we measured the improvement of the model. We calculated novel impressions ratio for each model, and the percentage increase of the novel impression ratio compared to the reference. The novel impression ratio is the percentage of model impressions that are solely recommended by this model. We see that the Multimodal LLM model proposed in this paper achieves the highest novelty gain. 
On the other hand, we calculated the positive feedback ratio for each model and compared it with the reference. It shows that our proposed model has better quality than the exploration-based model.

\begin{figure}[t]
    \centering
    \subfigure[\#Users watch >= 10min]
    {
    \includegraphics[width=0.45\columnwidth]{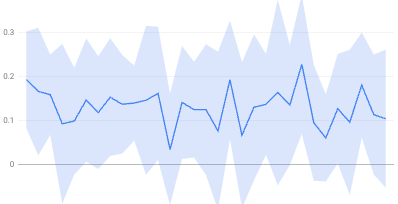}
    }
    \hspace{0.05in}
    \subfigure[Unique (User, Watched Item) Pairs]
    {
    \includegraphics[width=0.45\columnwidth]{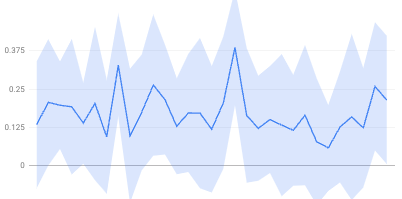}
    }
    \caption{The proposed method drives user growth. Y-axis shows the degree of improvement of live metrics.}
    \label{fig:live}
\end{figure}

\smallskip
\noindent\textbf{Live Metrics.} 
In Figure \ref{fig:live}, we measured the impact on users satisfaction by adding the proposed model to existing production models. In Figure \ref{fig:live}(a), the treatment has increased the number of daily users with more than 10 minutes engagement significantly. In Figure \ref{fig:live}(b), the treatment has improved the unique number of user and engaged item pairs too. This indicates that the proposed model improves the users satisfaction by providing serendipitous recommendations.

\begin{figure}[h!]
    \centering
    \includegraphics[width=0.3\textwidth]{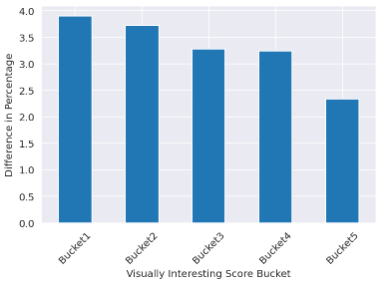}
    \caption{Users Satisfaction Gain of LLM nominated Candidates Across Different Context Videos Buckets}
    \label{fig:VisuallyInteresting}
\end{figure}

\noindent\textbf{Engagement analysis on different types of context videos} 
One of our hypotheses is that LLM-generated video candidates should perform better for context videos that have richer visual content. To analyze this, we used a visually interesting score to assess how rich the video's visual content is.  The visually interesting score is generated by a binary classifier trained on video content embeddings. The classifier generated raw score between 0 and 1 as the probability that the video is visually interesting. We grouped the context videos by their visually interesting score bucket, as shown in Figure \ref{fig:VisuallyInteresting}(b). Bucket1 videos are most visually interesting, while bucket 5 videos are least visually interesting. Then we calculated user value for LLM-nominated and non-LLM-nominated video candidates within each group. Utilizing the engagement rate of non-LLM-nominated candidates as a baseline, we observed the relative improvement in engagement rate from LLM-nominated candidates across the whole bucket. Moreover, our analysis showed a positive correlation: the more visually interesting the context videos are, the larger the user value gain achieved from LLM-nominated videos.

\section{Conclusion}
We introduce a scalable, hierarchical framework that uses a fine-tuned MLLM to steer a conventional recommender towards more serendipitous suggestions. Through live experiments, we demonstrate that a chain-of-thought approach with multimodal inputs significantly improves recommendation novelty and user satisfaction, providing a pioneering blueprint for applying MLLMs to enhance user exploration in real-world systems.

\balance
\bibliographystyle{ACM-Reference-Format}
\bibliography{sample_bib}

\end{document}